\documentclass[aoas,MSNbibl,nameyear,seceqn,dvips]{arximspdf}
\usepackage{mathrsfs}
\usepackage{graphicx}

\doi{10.1214/11-AOAS534}
\volume{6}
\issue{3}
\pubyear{2012}
\firstpage{1236}
\lastpage{1255}

\makeatletter
\renewcommand{\epsilon}{\varepsilon}
\makeatother

\begin{document}
\begin{frontmatter}

\title{Semiparametric zero-inflated modeling
in multi-ethnic study of atherosclerosis (MESA)}
\runtitle{Semiparametric zero-inflated modeling in MESA}

\begin{aug}
\author[A]{\fnms{Hai} \snm{Liu}\corref{}\ead[label=e1]{liuhai@iupui.edu}},
\author[B]{\fnms{Shuangge} \snm{Ma}\thanksref{t1}\ead[label=e2]{shuangge.ma@yale.edu}},
\author[C]{\fnms{Richard} \snm{Kronmal}\thanksref{t2}\ead[label=e3]{kronmal@u.washington.edu}}\\
\and
\author[D]{\fnms{Kung-Sik} \snm{Chan}\thanksref{t3}\ead[label=e4]{kung-sik-chan@uiowa.edu}}
\runauthor{Liu, Ma, Kronmal and Chan}
\affiliation{Indiana University School of Medicine, Yale University,
University~of~Washington and University of Iowa}
\address[A]{H. Liu\\
Department of Biostatistics\\
Indiana University School of Medicine\\
Indianapolis, Indiana 46202\\
 USA\\
\printead{e1}} 
\address[B]{S. Ma\\
School of Public Health\\
Yale University\\
New Haven, Connecticut 06520\\
 USA\\
\printead{e2}}
\address[C]{R. Kronmal\\
Department of Biostatistics\\
University of Washington\\
Seattle, Washington 98101\\
 USA\\
\printead{e3}}
\address[D]{K. S. Chan\\
Department of Statistics\\ \quad  and Actuarial Science\\
University of Iowa\\
Iowa City 52242, Iowa\\
 USA\\
\printead{e4}}

\end{aug}
\thankstext{t1}{Supported in part by US National Science Foundation Grant DMS-08-05984.}
\thankstext{t2}{Supported by Grant N01-HC95159 from the National Heart, Lung and Blood Institute.}
\thankstext{t3}{Supported in part by US National Science Foundation Grant DMS-09-34617.}

\received{\smonth{4} \syear{2011}}
\revised{\smonth{10} \syear{2011}}

\begin{abstract}
We analyze the Agatston score of coronary artery calcium (CAC) from
the Multi-Ethnic Study of Atherosclerosis (MESA) using the
semiparametric zero-inflated modeling approach, where the observed CAC
scores from this cohort consist of high frequency of zeroes and continuously
distributed positive values. Both partially constrained and unconstrained models are considered to investigate the
underlying biological processes of CAC development from zero to positive,
and from small amount to large amount. Different from existing studies,
a model selection procedure based on likelihood cross-validation is adopted
to identify the optimal model, which is justified by comparative Monte Carlo studies.
A shrinkaged version of cubic regression spline is used for model estimation and variable selection simultaneously.
When applying the proposed methods to the MESA data analysis, we show that the two biological
mechanisms influencing the initiation of CAC and the magnitude of CAC when it is positive are better characterized by an unconstrained
zero-inflated normal model. Our results are significantly different from those in published studies,
and may provide further insights into the biological mechanisms underlying CAC development in humans.
This highly flexible statistical framework can be applied to zero-inflated data analyses in other areas.
\end{abstract}


\end{frontmatter}

\setcounter{footnote}{3}
\section{Introduction}
\label{sec:intro}

The Multi-Ethnic Study of Atherosclerosis (MESA)\break [\citet{MESA2002}] is an ongoing longitudinal study of subclinical
cardiovascular disease (CVD) involving a cohort of more than 6500
men and women from six communities in the United States\vadjust{\goodbreak}
(\url{http://www.mesa-nhlbi.org/}). It was initiated by the National
Heart, Lung and Blood Institute in July 2000 to investigate the
prevalence, risk factors and progression of subclinical CVD in a
population-based multi-ethnic cohort. Agatston score
[\citet{Agatston1990}], which measures the amount of coronary artery
calcium (CAC), is an important predictor of future coronary heart
disease events [\citet{Min2010}; \citet{Polonsky2010}]. However, many healthy people
may have no detectable CAC; consequently, CAC equals zero with
substantial relative frequency, but otherwise it is a continuous
positive variable. That CAC has a~mixture distribution with an atom
at zero hampers its analysis by standard statistical methods. Such
data are referred to as ``zero-inflated'' and require the
development of more complex statistical models.

Zero-inflated data actually abound in many areas, for example, in
health care cost studies [\citet{Blough1999}], environmental science
[\citet{Agarwal2002}], ecological applications [\citet{Liuetal2010}],
etc. Among various models for analyzing data with excess zeroes,
the hurdle model [\citet{Mullahy1986}] has been proposed to handle both zero-inflation and zero-deflation
in count data, which consists of two parts: one binary model to determine whether
the response outcome is zero or positive and a second part conditional on the positive
responses if the ``hurdle is crossed.'' On the other hand,
a zero-inflated model [\citet{Lambert1992}] that assumes an underlying mixture distribution of probability mass at
zero and some continuous or discrete distribution (e.g., normal,
Poisson) has been widely used to analyze zero-inflated continuous data [\citet{Couturier2010}].
Note that both the hurdle model and the zero-inflated model are essentially equivalent
to the two-part model [\citet{Kronmal2005}; \citet{Zhou2006}]
when dealing with zero-inflated continuous data as the CAC score in MESA
[see \citet{Min2005} for discussion on comparing existing models for zero-inflated count data].
Therefore, we will not distinguish the aforementioned two models
and refer to the approach as the zero-inflated model in the following discussion.
Also note that the two-part model for zero-inflated continuous data, with the probit link for the binary model part,
is a special case of the Heckman model
[also known as Type II Tobit model, see \citet{Heckman1979}; \citet{Amemiya1984}].
Most existing zero-inflated models are in the parametric
setting, assuming that the covariate effects are linear (on proper
link scales). However, the assumption of linearity may not hold in
public health or medical research. Instead, the semiparametric
regression model [\citet{Ruppert2003}] provides a powerful tool to describing nonlinear
relationships between the covariates and response variables in such
situations. For instance, \citet{Lam2006} used the sieve estimator
to analyze zero-inflated count data from a public health survey.

In zero-inflated data analyses, it is often of interest to examine whether
the zero and nonzero responses are generated by related mechanisms.
In MESA,\vadjust{\goodbreak} it may provide useful insights into the biological
process on whether or not the risk factors of CVD influence the
probability of having positive CAC and the progression of CAC when
it is present in a similar way, which could be statistically
verified by introducing proportional constraints into the
zero-inflated model. Such a constrained zero-inflated model can be interpreted
by a latent biological mechanism involving an unobservable random threshold
and has been studied mostly in a parametric
framework. For example, \citet{Han2006} considered proportional constraints in two-part models
in MESA to promote better understanding of the mechanism that drives
the zero-inflation in CAC, and to estimate the model parameters more accurately (intuitively because fewer
parameters need to be estimated in a~constrained model)
as well. However, they did not take into
account the nonlinear relationships between some covariates
and the response variable in MESA [\citet{McClelland2006}].
\citet{Maetal2010} incorporated proportional constraints in a
semiparametric zero-inflated normal model when analyzing the same data set, but they only considered
a~universal proportionality parameter on all covariates, which is not flexible enough to handle
more complicated zero-inflation processes (see Section~\ref{sec:methods} for more discussion).
Therefore, it becomes necessary to study a more flexible partially constrained semiparametric
zero-inflated model to overcome the limitations of the existing investigations.
We note that similar techniques of imposing
proportional constraints on two sets of regression coefficients in
complex models were investigated by \citet{Albert1997}, \citet{Moulton2002}, among others.

In this paper, we propose a partially constrained semiparametric zero-inflated model to
analyze the CAC score in MESA, which provides a highly
flexible approach for delineating the zero-inflated data generating
process. Under the general partially constrained model framework, the unconstrained and constrained
zero-inflated models together make it possible to shed new light on the
relationship between the zero and nonzero data generating
processes, and the latter promotes estimation efficiency when the
postulated constraint holds. Cubic regression spline with shrinkage is adopted to estimate
nonparametric regression functions and to select important variables simultaneously.
Because of the complex model specification with a mixture distribution,
a model selection procedure based on cross-validated likelihood is implemented to
examine the prediction performance of the fitted models, and to choose the optimal
zero-inflated model from multiple candidate
models with various partial proportional constraints,
which avoids the problem of multiple testing by treating each
candidate model on an equal basis. Estimation of the proposed
zero-inflated model and statistical inference will also be discussed.
The outline of this paper is as follows. We introduce the semiparametric
zero-inflated model methodology in Section~\ref{sec:methods}. Simulation studies are
carried out to illustrate the proposed model estimation and selection methods in Section~\ref{sec:sim}.
The analytical results of the MESA data analysis are presented in Section~\ref{sec:mesa}. Some
concluding remarks are discussed in Section~\ref{sec:discussion}.

\section{Methods}
\label{sec:methods}

\subsection{Semiparametric zero-inflated model}

Statistical analysis of zero-\break inflated data cannot proceed under the
assumption of regular probability distribution due to the high
frequency of zeroes. If the nonzero responses are continuously
distributed, a zero-inflated normal (ZIN) model can be utilized, which
assumes a~mixture distribution of probability mass at zero and
a~normal distribution, after suitable transformation. Suppose that
given the covariate vectors $\mathbf{Z}=(Z_1,\ldots,Z_m)'$ and
$\mathbf{X}=(X_1,\ldots,X_k)'$, the conditional distribution of
the response variable $Y$ is zero-inflated normal:
\begin{equation}
Y | \mathbf{Z},\mathbf{X} \sim  \cases{
    0, & \quad   with probability  $(1-p)$,\cr
    \mathcal{N}(\mu,\sigma^2), & \quad  with probability $p$,
}
  \label{eq:mix}
\end{equation}
where the covariate effects of $\mathbf{Z}$ are parametric and
those of $\mathbf{X}$ are nonparametric. The
above ZIN model consists of two parts:
\begin{equation}
g(p)=\beta_0+\boldsymbol{\beta}'\mathbf{Z}+\sum_{i=1}^k h_i(X_i)
\label{eq:p}
\end{equation}
links the nonzero-inflation probability $p$ to the covariates via a
link function~$g$ (e.g., logit or probit function) in the binary part, and the linear part
\begin{equation}
\mu=\gamma_0+\boldsymbol{\gamma}'\mathbf{Z}+\sum_{i=1}^k s_i(X_i)
\label{eq:mu}
\end{equation}
describes the covariate effects on the normal mean response $\mu$.
In the semiparametric setting, $\beta_0$ and $\gamma_0$ are two intercept terms,
the regression coefficients $\boldsymbol{\beta}=(\beta_1,\ldots,\beta_m)'$ and
$\boldsymbol{\gamma}=(\gamma_1,\ldots,\gamma_m)'$ correspond to the
parametric effects in the two parts, respectively, and $h_i, s_i,
i=1,\ldots, k$, are two sets of nonparametric smooth functions.
By setting some parametric coefficients and/or some smooth functions to be identically zero,
equations (\ref{eq:p}) and~(\ref{eq:mu}) subsume the case that the two parts of the model involve different sets of covariates.
Each univariate smooth function $h_i(X_i)$ or $s_i(X_i),
i=1,\ldots, k$, can be estimated nonparametrically using a cubic regression spline,
which can be readily extended to a high-dimensional smoother using thin
plate spline [\citet{Wood2003}] to accommodate interaction between several continuous predictor variables.

Equations (\ref{eq:mix}), (\ref{eq:p}) and (\ref{eq:mu}) formulate
an unconstrained semiparametric ZIN model, which assumes that the
covariate effects on the probability of having a nonzero response and
the magnitude of the nonzero response may follow different data\vadjust{\goodbreak}
generating mechanisms. However, an interesting research question
arises as to whether the two processes are related to some extent
such that some covariates influence the two processes similarly. The
partially constrained zero-inflated modeling approach
[\citet{Liu-Chan-2010}] could be used to test the above hypothesis,
which assumes that some of the smooth components (operating on the
same covariates) in (\ref{eq:p}) and (\ref{eq:mu}) bear
proportional relationships with the constraints:
\begin{equation}
h_i = \delta_i s_i,    \qquad  i\in \mathscr{C}\subseteq \{1,\ldots,k\}, \label{eq:constraint}
\end{equation}
where $\mathscr{C}$ is the index set of the constrained smooth components; $\delta_i, i\in \mathscr{C}$, are unknown proportionality parameters. The covariates corresponding to those smooth
functions with proportional constraints then affect the nonzero-inflation probability
and the mean nonzero response proportionally on
the link scales. However, the other covariates with indices not in $\mathscr{C}$ may have different
impacts on the above two processes, which can be flexibly modeled by
the unconstrained components. Note that the unconstrained zero-inflated model is a special case
in the general partially constrained model framework with $\mathscr{C}=\varnothing$.

We consider proportional constraints in the zero-inflated model not only because they may result in more
parsimonious models, but also because they may admit biological interpretation connected to some latent threshold model.
To illustrate this connection, suppose that $Y^*$ is a latent response variable following the
$\mathcal{N}(\mu,\sigma^2)$ distribution. The observed response~$Y$ is zero if the latent mean response $\mu$ is less
than a random threshold $T$ which could be due to measurement error or limits of detection,
and it is equal to $Y^*$ if $\mu$ exceeds the threshold. Hence, the nonzero-inflation probability
$p=\operatorname{Pr}(Y=Y^*)=\operatorname{Pr}(T\leq \mu)=F_{T}(\mu)$, where $F_{T}$ is the cumulative distribution function (CDF) of the random threshold variable $T$.
As a result, we would have $g(p)=\mu$ if the link function is taken as the inverse CDF of $T$, which is, however,
generally unknown. Nevertheless, according to \citet{LiDuan1989}, under some mild
regularity conditions, any maximum likelihood-type estimator is consistent up to a multiplicative scalar,
even under a misspecified link function. More specifically, if we use, for example, a logit link in (\ref{eq:p}),
the parameter estimators in the binary part are proportional to the true parameters in (\ref{eq:mu}),
that is, $\widehat{\boldsymbol{\beta}}=\delta \boldsymbol{\gamma}$, and $\hat{h}_i = \delta s_i$, $i=1,\ldots, k$, for some scalar $\delta$.
Alternatively, assuming the zero-inflation is caused by
some other biological characteristic depending on the covariates through $\xi(\mathbf{Z},\mathbf{X})$,
that is, $Y=0$ if $\xi(\mathbf{Z},\mathbf{X})< T$, we may have partial proportionality among the parameters in the two parts.
Based on this latent biological process, we further relax the proportionality parameter to be possibly different
across the linear and smooth components, leading to the proposed partially proportionally constrained zero-inflated model.

As a closely relevant study in the literature, \citet{Maetal2010}
compared the unconstrained semiparametric zero-inflated model to a fully proportionally
constrained model,\vadjust{\goodbreak} which assumed (\ref{eq:p}) and
$\mu=\alpha+\tau \{\boldsymbol{\beta}'\mathbf{Z}+\sum_{i=1}^k h_i(X_i) \}$,
with $\tau$ being the universal proportionality scale parameter.
The fully constrained model is, however, quite inflexible and it cannot
handle the cases with nonidentical sets of covariates in (\ref{eq:p}) and (\ref{eq:mu})
or more complicated zero-inflation mechanisms [e.g., $\xi(\mathbf{Z},\mathbf{X})\neq \mu$ as discussed above].
The \emph{partially constrained semiparametric zero-inflated model}, on the other hand,
is more flexible by untangling the constrained and unconstrained
smooth components. In addition, when the postulated proportional constraint
holds, the more parsimonious partially constrained zero-inflated
model promotes estimation efficiency compared to its unconstrained counterpart.
Compared to a more recent study by \citet{Maetal2011} on similar problems in a constrained semiparametric two-part model,
our method is computationally more affordable and more flexible.
In this study, we shall focus on the statistical inference and
model selection regarding proportional constraints on the
nonparametric smooth components, which has not been discussed in the literature to our knowledge.
Moreover, the estimation and inference methods proposed below could be
readily lifted to the cases where the parametric terms are also (partially) constrained.

\subsection{Model estimation and inference}

The proposed semiparametric zero-inflated model can be estimated by the
penalized likelihood approach, which, in the
unconstrained case, maximizes the following penalized log-likelihood function:
\[
\mathbb{P}_n
\ell(\beta_0,\boldsymbol{\beta},\gamma_0,\boldsymbol{\gamma},\sigma^2,h_1,\ldots,h_k,
s_1, \ldots, s_k) - \sum_{i=1}^{k}\lambda^2_{n,i}J(h_{i}) -
\sum_{i=1}^{k}\varphi^2_{n,i}J(s_{i}),
\]
where $\mathbb{P}_n$ is the empirical measure of $n$ observations,
$\ell=I(Y=0)\log(1-p)+I(Y\neq 0) \{\log
p-\frac{(Y-\mu)^2}{2\sigma^2} \}$ is the log-likelihood
function for a single observation, $J(f)$ defines a roughness
penalty functional of $f$, and $\lambda_{n,i},\varphi_{n,i},
i=1,\ldots,k$, are the smoothing parameters corresponding to each
penalty term, which control the trade-off between the smoothness of
the function estimates and goodness of fit of the model.
In this study, cubic regression spline is adopted with
roughness penalty $J(f)=\int \{f^{(2)}(x)\}^2\,dx$, where
$f^{(2)}(x)$ denotes the second derivative of a univariate function $f(x)$.
The spline estimate can be represented as a linear combination of some basis functions:
$\hat{f}(x)=\theta_0+\theta_1 x + \sum_{j=1}^{K-1} \theta_{j+1}(x-x_j^*)_+^3$, where
$x_j^*, j=1,\ldots,K-1$, are\vspace*{1pt} fixed knots placed evenly (in terms of percentiles) over the corresponding observed covariate values
[see \citet{Durrleman1989} for more discussion on the knots selection in cubic splines],
$(x)_+=x$ if $x>0$ and $(x)_+=0$ otherwise,
$\boldsymbol{\theta}=(\theta_0,\ldots,\theta_K)'$ is the parameter vector.
Accordingly, the roughness penalty could
be written as a quadratic form of the corresponding parameters, such that
$J(f)=\boldsymbol{\theta}'\mathbf{S} \boldsymbol{\theta}$, where $\mathbf{S}$ is the penalty matrix.
The smoothing parameters can be selected by generalized cross-validation (GCV) or similar
procedures. Under the main regularity conditions as following: (R1) The covariates $\{\mathbf{Z},\mathbf{X}\}$ and the true parametric coefficients $\beta_0$,
$\boldsymbol{\beta}$, $\gamma_0$, $\boldsymbol{\gamma}$ are bounded;
(R2)~$h_{i}, s_{i}$ are nonconstant and satisfy $J(h_{i}), J(s_{i})<\infty$; (R3)~$\lambda_{n,i}, \varphi_{n,i}=O_{\mathbf{P}}(n^{-2/5})$;
(R4)~The Fisher information matrix is nonsingular,
plus some minor technical conditions, the maximum penalized likelihood estimators of the smooth functions
can be shown to be $n^{2/5}$ consistent, and the parametric coefficient estimators are
$n^{1/2}$ consistent and asymptotically normal, using similar empirical processes techniques in \citet{Liu-Chan-2010}.

Statistical inference including construction of the confidence intervals for the parametric coefficients
and confidence bands for the smooth functions can be based on the observed Fisher information matrix,
which avoids computer-intensive bootstrap methods used in \citet{Maetal2010} and \citet{Maetal2011}.
Monte Carlo studies reported in \citet{Liu-Chan-2010}
showed that such confidence intervals/bands enjoyed desirable empirical properties in that their
across-the-function coverage rates were close to their nominal levels.
Estimation and inference of a partially constrained semiparametric
zero-inflated model follow a similar procedure.
More details of the estimation algorithm and
theoretical results can be found in \citet{Liu-Chan-2010}.

As pointed out by \citet{Wood2006}, a disadvantage of the cubic spline smoother is that the estimated smooth is never
completely eliminated in the sense of having all corresponding parameters estimated to be zero.
In addition, linear components in the smooth function are always unpenalized by the second derivative penalty.
From a variable selection point of view [\citet{Huang2010}], it may be desirable to have the smooth  be shrunk completely to
zero if the corresponding smoothing parameter is sufficiently large, and preserve the curvature otherwise.
\citet{Wood2006} proposed to add an extra small amount of ridge-type
of penalty to the original penalty matrix, that is, $\mathbf{S}_\epsilon = \mathbf{S} + \epsilon\mathbf{I}$
was used as the penalty matrix with additional shrinkage. The parameters of a smooth function with large smoothing
parameter are set to be exactly zero. But otherwise the additional small fraction of an identity matrix has almost no influence on the
cubic spline estimate if it is not shrunk to linearity by the roughness penalty.
With this slight adjustment, the resulting cubic smooth
with additional shrinkage behaves reasonably well in variable selection empirically,
which is illustrated in a simulation study in Section~\ref{sec:sim}.
In the following discussion, the cubic regression spline with shrinkage is adopted to estimate the
nonparametric covariate effects as well as to select relevant variables simultaneously.

\subsection{Partial-constraint selection}

One remaining issue with the partially constrained zero-inflated model
is to choose an optimal model in terms of prediction performance from multiple candidate models with
various partial constraints [model (\ref{eq:mix}) to (\ref{eq:mu}) with constraint
(\ref{eq:constraint}), note that different index sets $\mathscr{C}$ correspond
to\vadjust{\goodbreak}
different partially constrained models, including $\mathscr{C}=\varnothing$, that is, the unconstrained model]
and justify the selection procedure.
\citet{Liu-Chan-2010} proposed a model selection criterion for a nonparametric zero-inflated model based on
the marginal likelihood, which is similar to the Bayesian information criterion (BIC) [\citet{Schwarz1978}].
However, although the marginal likelihood criterion was shown to work
well for zero-inflated model selection both theoretically and empirically,
it was derived for penalized regression splines without additional
shrinkage. Little is known about its behavior when applied to the shrinkaged version of cubic spline, as we adopted in
this study. Instead, cross-validation works almost universally [\citet{Shao1993}]
for most model selection purposes, which assesses the prediction performance of the models under comparison.
The model selection method is easier to
implement in practice than the hypothesis testing approach used in other studies [see, e.g., \citet{Han2006}],
which usually involves step-wise search and whose complexity increases
dramatically with the number of candidate models.

Among a variety of cross-validation methodologies [\citet{Arlot2010}],
we use the Monte Carlo cross-validation (MCCV) [\citet{Picard1984}] to examine the out-of-sample prediction
performances of various partially constrained zero-inflated models under consideration. In particular,
the data are randomly partitioned into two disjoint sets,
one of which with a fixed fraction $1-\nu$ of the whole data (training set)
are used to build the model, and the remaining $\nu$ fraction of the data
(validation set) are used to evaluate some goodness-of-fit criterion (or, equivalently, the risk)
for each candidate model. The partition is repeated independently for $B$ times
and the out-of-sample prediction performance of each model is estimated by taking
the average over the $B$ validation sets.
Furthermore, because of the complexity of the mixture zero-inflated distribution, the goodness-of-fit criterion
need to be chosen with caution. We propose to use cross-validated likelihood as the prediction performance
criterion, which is advocated in a probabilistic clustering problem using mixture modeling [\citet{Smyth2000}].
The cross-validated (log-)likelihood of the $k$th candidate model is defined as
\[
\ell_{k}^{\mathrm{cv}}=\frac{1}{B}\sum_{j=1}^{B}\ell \bigl(\widehat{\Theta}_k(D\setminus D_j^{\mathrm{v}}) |D_j^{\mathrm{v}} \bigr),
\]
where $D$ denotes the original data, $D_j^{\mathrm{v}}$ is the validation set of the $j$th partition,
$\widehat{\Theta}_k(D\setminus D_j^{\mathrm{v}})$ is the maximum penalized likelihood estimator
of the model parameter for the $k$th candidate model estimated from the $j$th training set,
and $\ell$ is the (log-)likelihood function evaluated on $D_j^{\mathrm{v}}$. It can be
shown that the expected value of the likelihood evaluated on an independent validation data set
is related to the Kullback--Leibler divergence between the truth and the model under consideration [\citet{Smyth2000}].

Other possible model selection criteria include the mean squared error (MSE) of the\vadjust{\goodbreak}
nonzero responses, and the area under the receiver operating
characteristic (ROC) curve [AUC, larger is preferred as it indicates better prediction,
see, e.g., \citet{Miller1991}] of the binary indicators of zero responses. However, both MSE and AUC have
limitations when applied to zero-inflated data. In particular, the MSE only measures the risk for the nonzero responses, whereas the AUC takes into account all validation samples, but it fails to assess the accuracy of the
predictive value of the nonzero response. Sometimes the two criteria may point to different candidate models,
which confounds the model selection. In addition, the bias-corrected MSE
[denoted as $\mathrm{MSE}_c$, it is not difficult to see that $\textsf{E}(Y)=p\mu$ from (\ref{eq:mix})]
of both the zero and nonzero data can be calculated for each validation set as
\[
\mathrm{MSE}_c=\frac{1}{n_{\mathrm{v}}}\sum_{i=1}^{n_{\mathrm{v}}}{ (\hat{p}_i \hat{\mu}_i-y^{\mathrm{v}}_i )^2},
\]
where $y^{\mathrm{v}}_i$ is the $i$th observed response in the validation
set with $n_{\mathrm{v}}$ samples, $\hat{p}_i$ and $\hat{\mu}_i$ are the corresponding estimated
nonzero-inflation probability and mean
nonzero-inflated response from the fitted model, respectively.
A simulation study is carried out to evaluate the performance in model selection between
the partially constrained and unconstrained zero-inflated models based on the cross-validated likelihood,
AUC, MSE and MSE$_c$ in Section~\ref{sec:sim}. In practice, it suffices to set the MCCV replication size
$B$ to some number between 20 and 50 in most model selection problems in the parametric framework [\citet{Shao1995}].
In the semiparametric setting, as in our study, we repeat the partition with equal size ($\nu=0.5$) for $B=100$ times in
the simulation study and $B=200$ in the MESA data analysis because of its much larger sample size.

\section{Simulation study}
\label{sec:sim}

Before applying the zero-inflated modeling approach to the MESA data
analysis, we first conduct a Monte Carlo study to examine the performance
of the proposed model selection method based on the cross-validated likelihood, as well
as other goodness-of-fit criteria discussed in the previous section.
The out-of-sample prediction performance is evaluated for two candidate models, that is,
the partially constrained zero-inflated normal model (with the correct model specification)
and its unconstrained counterpart.

\begin{figure}[b]

\includegraphics{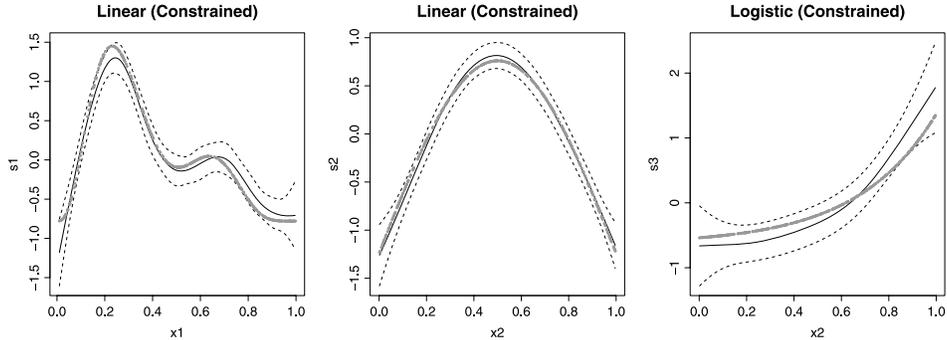}

  \caption{Estimated smooth functions fitted by the partially
  constrained zero-inflated normal model, with $n=400$ and $\sigma=0.5$.
  The solid lines show the cubic regression spline estimates, with the dashed
  lines representing the 95\% point-wise confidence bands.
  The gray dots denote the true functional values.}
  \label{fig:sim1}
\end{figure}

The simulated data were generated
based on three univariate test functions $s_1$, $s_2$ and $s_3$ on $[0,1]$:
\begin{eqnarray*}
s_{1}(x)&=& \bigl\{0.2x^{11}\bigl(10(1-x)\bigr)^{6}+10(10x)^{3}(1-x)^{10} \bigr\}/4, \\
s_{2}(x)&=&2\sin(\pi x), \\
s_{3}(x)&=&\exp(3x)/10.
\end{eqnarray*}
First, $n$ independently uniformly distributed random variables $X_1,
X_2$ and~$X_3$ were generated on $[0,1]$. A two-level factor covariate
$Z$ was set to be~0\vadjust{\goodbreak} for the first $n/2$ samples and 1 for the rest.
The true nonzero-inflation probability~$p$ and nonzero-inflated
mean response $\mu$ were generated by
\begin{eqnarray}\label{eq:sim1}
\mathrm{logit}(p) &=&0.3Z+0.5\bar{s}_1(X_1)+\bar{s}_2(X_2),  \\\label{eq:sim2}
\mu &=&-1+2Z+\bar{s}_1(X_1)+\bar{s}_3(X_2),
\end{eqnarray}
where each smooth component was centered at the observed covariate
values and denoted as $\bar{s}_j, j=1,2,3$, respectively. The
nonzero-inflated responses $Y^*_i, i=1,\ldots,n$, were randomly
sampled from normal $\mathcal{N}(\mu_i, \sigma)$ distributions. The
response variable was then ``zero-inflated'' according to the
indicator random variables $E_i$, which followed independent $\operatorname{Bernoulli}(p_i)$ distributions, that is, $Y_i=Y^*_i$ if $E_i=1$ and
$Y_i=0$ if $E_i=0$. The simulated data set is denoted as
$\{(Y_i,Z_i,X_{i1},X_{i2},X_{i3})\}_{i=1}^n$. Note that the above data
simulation procedure specifies a partially constrained ZIN with
proportional constraint on the $\bar{s}_1(X_1)$ component. But the covariate
$X_2$ affects the probability of nonzero inflation and the mean
nonzero response with different functional forms, namely, $s_2$ and
$s_3$ in equations (\ref{eq:sim1}) and (\ref{eq:sim2}), respectively.
$X_3$ is a~redundant covariate that has no impact in either $p$ or $\mu$.

\begin{figure}

\includegraphics{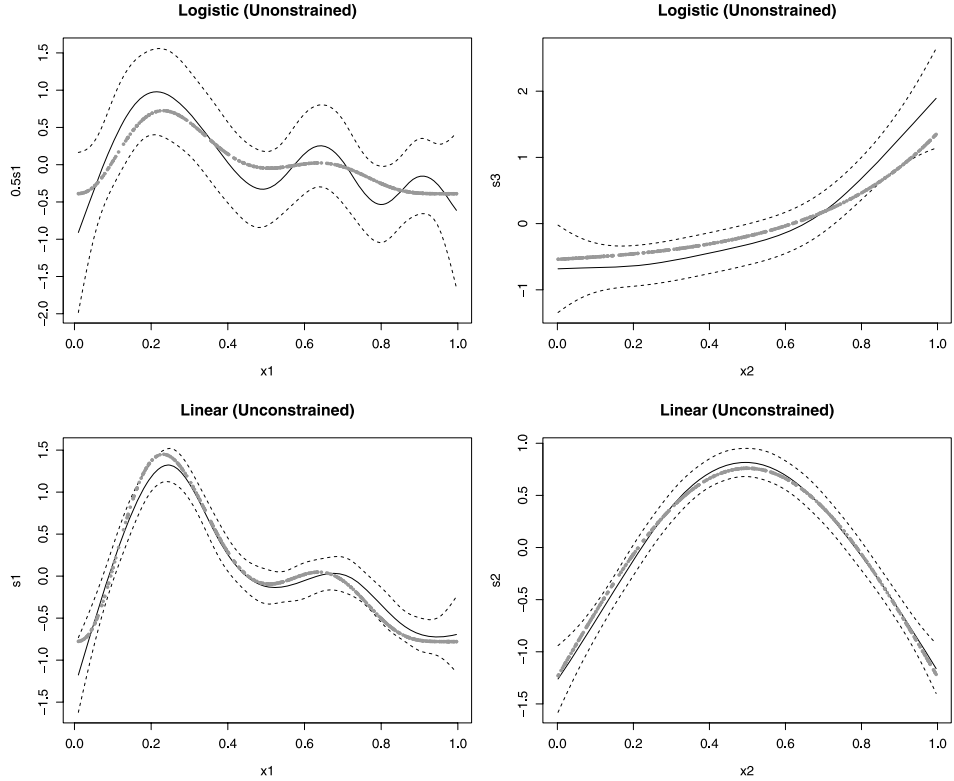}

  \caption{Estimated smooth functions with 95\% point-wise confidence bands (dashed
  lines) fitted by the unconstrained zero-inflated normal model, with $n=400$ and $\sigma=0.5$.
  The gray dots denote the true functional values.}
  \label{fig:sim2}
  \vspace*{-2pt}
\end{figure}

For each simulated data set, we fitted the partially constrained ZIN
and the unconstrained counterpart, with nine evenly spaced knots for each cubic spline.
We examined seven sample sizes from
$n=200$ to $800$, with two noise levels $\sigma=0.5$ and 1.
Figure~\ref{fig:sim1} shows the smooth function
estimates by the partially constrained ZIN fitted to one simulated data set with $n=400$ and
$\sigma=0.5$. The estimated smooth functions by the unconstrained ZIN fitted to
the same data set are displayed in Figure~\ref{fig:sim2}.
The wide confidence band (i.e., large standard errors of the smooth function estimates)
in the upper left panel of Figure~\ref{fig:sim2} suggests the lack of
efficiency in estimating the logistic part~(\ref{eq:sim1}) by the unconstrained ZIN as compared to the constrained
model, if the data were generated from the constrained model
[see \citet{Liu-Chan-2010}, Section 5, for more discussion on the estimation efficiency].
It is worthwhile to mention that in both the constrained and unconstrained models,
the covariate effect of the redundant variable $X_3$ was completely shrunk to zero
by the cubic regression splines with shrinkage.

\begin{figure}

\includegraphics{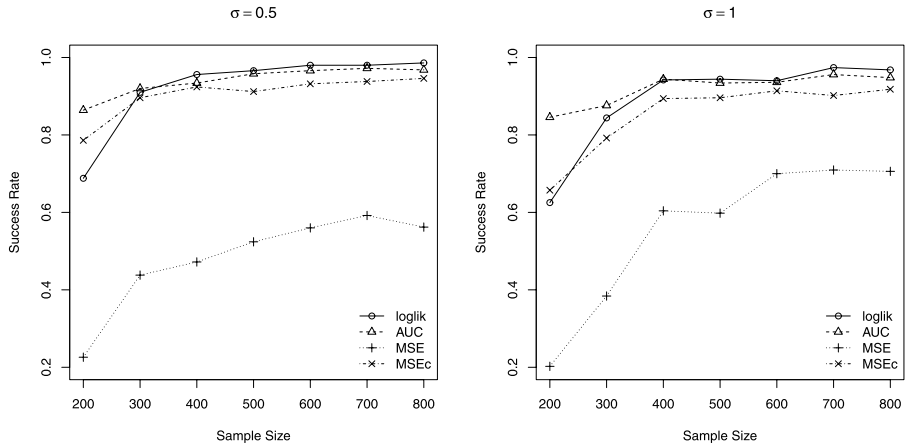}

  \caption{Model selection performance of the cross-validated likelihood, AUC, MSE and MSE$_c$.}
  \label{fig:simsum}
\end{figure}

MCCV were conducted for $B=100$ times with $\nu=0.5$ to evaluate the cross-validated
likelihood, AUC, MSE and MSE$_c$ for the constrained and unconstrained ZIN models.
In each of the aforementioned settings, 500 replications were
performed and the success rates in selecting the true model by each of the criteria
were compared and summarized in Figure~\ref{fig:simsum}.
As expected, the success rates of selecting the true model generally increase
with the sample size for each criterion. Except for very small sample sizes ($n=200, 300$, note that
they were zero-inflated data with nearly 50\% zeroes),
the cross-validated likelihood outperforms all other three criteria. Especially for
MSE of the nonzero data, the success rates are significantly lower than the other three
in each scenario,\vadjust{\goodbreak} which suggests that it is not a reliable measure of the overall prediction
performance for zero-inflated model. On the other hand, the bias-corrected version of MSE performs
reasonably well. By comparing the levels of error variance, the success rates are observed to be
consistently higher across different sample sizes for $\sigma=0.5$ (left panel of Figure~\ref{fig:simsum})
than $\sigma=1$ (right panel),
except for MSE. This seemingly implausible phenomena of MSE may be explained by the bias--variance trade-off
of the imposed constraint in the zero-inflated model. More specifically, the constrained model is
more parsimonious and hence has smaller estimation variance as compared to the unconstrained model,
which may also introduce bias. When the error variance is reduced, the bias becomes more dominant
than the estimation variance in the MSE decomposition.
Therefore, the unconstrained model tends to be more favored by
MSE when the error variance is smaller.

We also remark that the average discrepancies of all criteria between the
unconstrained and correctly specified constrained models decrease
as the sample size increases, which suggests that the relative predictive gain by the
constrained ZIN diminishes with increasing sample size. This is not surprising because as the sample size increases,
the estimation error becomes smaller relative to the intrinsic variabilities in the data.
So if the data were truly generated from a partially constrained zero-inflated
model and the sample size is large, we would benefit not as much on the estimation efficiency by fitting a
constrained model as for small to moderately large sample sizes.
In situations where there are too few observations to
carry out an unconstrained semiparametric or nonparametric
zero-inflated model analysis, fitting a partially constrained model
may provide an elegant alternative---a perspective earlier advanced
by \citet{Lambert1992} within the parametric zero-inflated framework.\vadjust{\goodbreak}

\begin{figure}[b]

\includegraphics{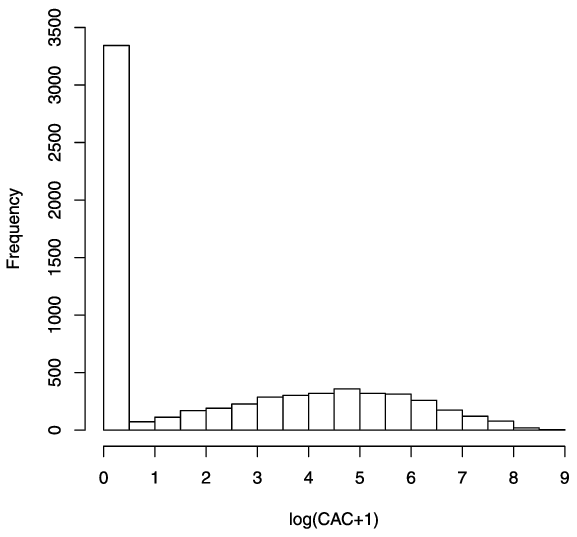}

  \caption{Histogram of $\log(\mathrm{CAC}+1)$ from MESA.}
  \label{fig:histy}
\end{figure}

In summary, as illustrated by the Monte Carlo study, the cross-validated likelihood performs very well in selecting
the true model with over 90\% success rate under mid to large sample sizes ($n\geq 400$),
which provides strong justification for the proposed model selection procedure.

\section{MESA data analysis}
\label{sec:mesa}

\subsection{Model specification}

The MESA data consist of 6672 participants 44 to 84 years old
(after removing missing values), among which 3343 (50.1\%) have
zero Agatston scores of CAC. We use $\log(\mathrm{CAC}+1)$ as the response
variable (the log-plus-one transformation is commonly used in
many applications to avoid long tails and preserve the zeroes),
and the covariates include gender (0-female, 1-male), race
(0-Caucasian, 1-Chinese, 2-African American, 3-Hispanic), diabetes
mellitus (0-normal, 1-otherwise), cigarette smoking status (0-never,
1-former, 2-current), age, body mass index (BMI), diastolic blood
pressure (DBP), systolic blood pressure (SBP), high-density
lipoprotein (HDL) cholesterol and low-density lipoprotein (LDL)
cholesterol, of which the first four could be treated as factor
predictors, and the rest are continuous variables. Because
approximately half of the CAC scores are zeroes, while the remaining
are positive and continuously distributed, we adopt a semiparametric
zero-inflated normal regression model for the response variable
$Y= \log(\mathrm{CAC}+1)$ (see Figure~\ref{fig:histy}),
with the conditional response distribution as
specified by (\ref{eq:mix}). The covariate
effect of BMI was found to be linear in a preliminary analysis, hence, it was modeled as
a~parametric term. There was only slight interaction between HDL and
LDL cholesterol levels, so they were modeled additively for ease of
interpretation. As a consequence, the probability of having positive
CAC is linked via the logit function (it is referred to as logistic
part and henceforth) to the covariates as follows:
\begin{eqnarray}
\label{mesa:p}
\mathrm{logit}(p)&=& \beta_0+\beta_1 \mathit{Male}+\beta_2 \mathit{Chinese} + \beta_3
\mathit{Black} + \beta_4 \mathit{Hispanic} \nonumber\\
&&{}+ \beta_5 \mathit{Cig}_f + \beta_6 \mathit{Cig}_c + \beta_7 \mathit{DM} + \beta_8 \mathit{BMI} \\
&&{}+ h_1(\mathit{Age})+h_2(\mathit{DBP})+h_3(\mathit{SBP})+h_4(\mathit{HDL})+h_5(\mathit{LDL}),\nonumber
\end{eqnarray}
where $\beta_0$ to $\beta_8$ are the regression coefficients
associated with the parametric terms; $\mathit{DM}$ stands for abnormal
diabetes mellitus status; $\mathit{Cig}_f$ and $\mathit{Cig}_c$ are binary indicators
of former and current smoker respectively; $h_i, i=1, \ldots, 5,$
are unknown smooth functions. The mean
positive (transformed) CAC level is specified as
follows (linear part):
\begin{eqnarray}
\label{mesa:mu}
\mu&=& \gamma_0+\gamma_1 \mathit{Male}+\gamma_2 \mathit{Chinese} + \gamma_3 \mathit{Black} +
\gamma_4 \mathit{Hispanic} \nonumber \\
&&{}+ \gamma_5 \mathit{Cig}_f + \gamma_6 \mathit{Cig}_c + \gamma_7 \mathit{DM} + \gamma_8 \mathit{BMI} \\
&&{}+ s_1(\mathit{Age})+s_2(\mathit{DBP})+s_3(\mathit{SBP})+s_4(\mathit{HDL})+s_5(\mathit{LDL}),\nonumber
\end{eqnarray}
where $\gamma_0$ to $\gamma_8$ are regression coefficients, $s_i,
i=1, \ldots, 5,$ are smooth functions possibly distinct from $h_i$.
All univariate smooth functions in the logistic and linear parts above
were estimated nonparametrically using cubic regression splines with shrinkage
and nine evenly spaced knots to identify important risk factors, as
discussed in Section~\ref{sec:methods}.

The unconstrained ZIN models (\ref{mesa:p}) and (\ref{mesa:mu})
assume that the covariate effects on the probability of having a
positive CAC score and the mean positive (transformed) CAC may be
driven by different physiological processes. As discussed earlier,
a partially constrained zero-inflated model could be used to test
whether the two processes are related to some extent. For example,
we can add a proportional constraint $h_1=\delta_1 s_1$ to examine
whether age acts in a similar manner in affecting CAC from zero to
positive, and from small amount to large amount. By comparing the
fitted partially constrained and unconstrained models based on their cross-validated likelihoods,
we can properly address interesting scientific hypotheses as above,
which may help elucidate the biological process responsible for CAC development.

Table~\ref{tab:cand} lists some partially constrained and
unconstrained ZIN models fitted to the MESA data, with corresponding
cross-validated (log-)likelihood, AUC, MSE and MSE$_c$ estimated from
$B=200$ replications (given that the sample size is considerably larger than
that in the simulation study) of MCCV with equal size of training set and validation set ($\nu=0.5$).
The DBP effect was found to be completely eliminated in both the logistic and linear parts and, hence,
it was treated as an unconstrained component.
We did not include models with constraints on the HDL
and/or LDL components due to convergence problem. This suggests that
the HDL and LDL effects are likely to be very different in the two
processes, namely, the absence/presence of CAC and the level of CAC
when it is present, such that forcing\vadjust{\goodbreak} them to be proportional on the
link scales will cause numerical problems in the estimation. Therefore,
we considered four candidate semiparametric zero-inflated models in the MESA data analysis:
$M_1$---no constraint imposed, $M_2$---proportional constraint on age,
$M_3$---proportional constraint on SBP, and
$M_4$---proportional constraints on both age and SBP (with different proportionality parameters).

\begin{table}
\tabcolsep=0pt
\caption{Comparison of candidate zero-inflated normal
models fitted to the MESA data based on Monte Carlo cross-validation with $B=200$ and $\nu=0.5$.
``$\checkmark$'' denotes the proportional
constrained smooth component; ``$\times$'' denotes the unconstrained
smooth component}\label{tab:cand}
\begin{tabular*}{\textwidth}{@{\extracolsep{\fill}}lccccccccc@{}}
  \hline
  \textbf{Model}&\textbf{Age}&\textbf{DBP}&\textbf{SBP}&\textbf{HDL}&\textbf{LDL}&\textbf{loglik}&\textbf{AUC}&\textbf{MSE}&\textbf{MSE$_{\boldsymbol c}$}\\
    \hline
  $M_1$ & $\times$ & $\times$ & $\times$ & $\times$ & $\times$ &  $-5018.5$  & 0.79 & 2.59 & 4.38\\
  $M_2$ & $\checkmark$ & $\times$ & $\times$ & $\times$ & $\times$ &  $-5034.9$  & 0.78 & 2.60 & 4.42\\
  $M_3$ & $\times$ & $\times$ & $\checkmark$ & $\times$ & $\times$ &  $-5022.3$  & 0.79 & 2.61 & 4.38\\
  $M_4$ & $\checkmark$ & $\times$ & $\checkmark$ & $\times$ & $\times$ &  $-5034.6$  & 0.78 & 2.60 & 4.42\\
  \hline
\end{tabular*}
\end{table}

According to the cross-validated likelihood, the
unconstrained ZIN mod\-el~$M_1$ has the best prediction performance among all candidate models.
The second best model is the constrained model $M_3$, which imposes
a proportional constraint on the SBP component (estimated
proportionality parameter is 0.682 with standard deviation
0.157). Note that all other three criteria, that is, AUC, MSE, and MSE$_c$, are very close, especially
for $M_1$ and $M_3$. This is expected because, as discussed in the end of Section~\ref{sec:sim},
the discrepancies between these criteria would be very small with large sample size.
In fact, the AUC and MSE$_c$ criteria (which are two reasonably reliable measures as
demonstrated in the simulation study) of $M_1$ and $M_3$ are so close that
it is hard to discern any differences. However, there is still
some gain in the cross-validated likelihood by fitting an unconstrained ZIN as compared to the
constrained models. We also tried other values of the fraction $\nu$
between 0.5 and 0.85 with more data in the validation set
($\nu=0.85$ corresponds to 1000 samples in the training set)
to assess the robustness of the likelihood cross-validation procedure.
The unconstrained model was consistently selected under various sizes of the validation set.
Therefore, according to the prediction performance using cross-validation, the unconstrained model performs
better than the partially constrained models, which suggests that
the covariates act differently in predicting the presence of
positive CAC and its severity when it is positive. The above result is significantly
different from existing studies, including \citet{Han2006}, \citet{Maetal2010} and \citet{Maetal2011} in
the determination of proportional constraint in zero-inflated models of CAC score in MESA.

\begin{table}
\tabcolsep=0pt
\caption{Coefficient estimates of the fitted
 unconstrained zero-inflated normal model defined by equations (\protect\ref{mesa:p}) and (\protect\ref{mesa:mu})}\label{tab:par}
\begin{tabular*}{\textwidth}{@{\extracolsep{\fill}}lcccccc@{}}
  \hline
   & \multicolumn{3}{c}{\textbf{Logistic}} & \multicolumn{3}{c@{}}{\textbf{Linear}} \\[-5pt]
   & \multicolumn{3}{c}{\hrulefill} & \multicolumn{3}{c@{}}{\hrulefill}
   \\
   & \textbf{Estimate}  & \textbf{SE}  & \textbf{\emph{p}-value} & \textbf{Estimate} & \textbf{SE} & \textbf{\emph{p}-value} \\
  \hline
  $\mathit{Intercept}$ &  $-1.13$ &  0.18 & ${<} 0.001$ & \hphantom{$-$}$3.46$ &  0.19 & ${<} 0.001$ \\
  $\mathit{Male}$   &   \hphantom{$-$}0.91 &  0.06 & ${<} 0.001$ &  \hphantom{$-$}0.67 & 0.06  & ${<} 0.001$ \\
  $\mathit{Chinese}$ &   $-0.13$ &  0.10 & \hphantom{${<}$}$0.209$ &  $-0.28$ & 0.10  & \hphantom{${<}$}0.004 \\
  $\mathit{African}$ &   $-0.79$ &  0.07 & ${<} 0.001$ &  $-0.37$ & 0.07  & ${<} 0.001$ \\
  $\mathit{Hispanic}$ &   $-0.63$ &  0.08 & ${<} 0.001$ & $-0.34$  & 0.08  & ${<} 0.001$ \\
  $\mathit{DM}$   &   \hphantom{$-$}0.25 &  0.07 & ${<} 0.001$ &  \hphantom{$-$}0.27 & 0.06  &  ${<} 0.001$ \\
  $\mathit{Cig}_f$  &   \hphantom{$-$}0.37 &  0.06 & ${<} 0.001$ & \hphantom{$-$}0.19  & 0.06  & \hphantom{${<}$}0.002 \\
  $\mathit{Cig}_c$  &   \hphantom{$-$}0.61 &  0.09 & ${<} 0.001$ & \hphantom{$-$}0.31  & 0.09  & ${<} 0.001$ \\
  $\mathit{BMI}$  &   \hphantom{$-$}0.03 &  0.01 & ${<} 0.001$ & \hphantom{$-$}0.02  & 0.01  & \hphantom{${<}$}0.002 \\
  \hline
\end{tabular*}
\end{table}

\subsection{Analytical results}

We now present the results of the fitted unconstrained
semiparametric zero-inflated model to the MESA data, as selected by the\vadjust{\goodbreak}
model selection procedure based on likelihood cross-validation.
Table~\ref{tab:par} lists the coefficient estimates of the
parametric components. The model results suggest that men have
increased risk of having positive CAC and higher mean CAC score when
it is present, as compared to women. Both African and Hispanic Americans have
reduced probability of having CAC, as compared with Caucasian.
Chinese, African and Hispanic Americans all have lower average CAC level when
it is positive. Having abnormal diabetes status will increase both
the risk of positive CAC and its progression. Former smokers are
more likely to have CAC and, on the average, they have higher
positive CAC scores, as compared with nonsmokers. Current smokers
have even higher risk and mean positive CAC level. BMI is linearly
positively associated with both the probability of having positive
CAC (on the link scale) and CAC score if it is positive.

Among other related MESA studies, \citet{Han2006} included only gender, race and age as the covariates,
and their parameter estimates are similar to ours in signs and magnitudes,
except that they found Chinese had significantly reduced risk of having positive CAC,
as compared to Caucasian. The above findings on the parametric components are consistent with the
unconstrained two-part model in \citet{Maetal2010}.

\begin{figure}

\includegraphics{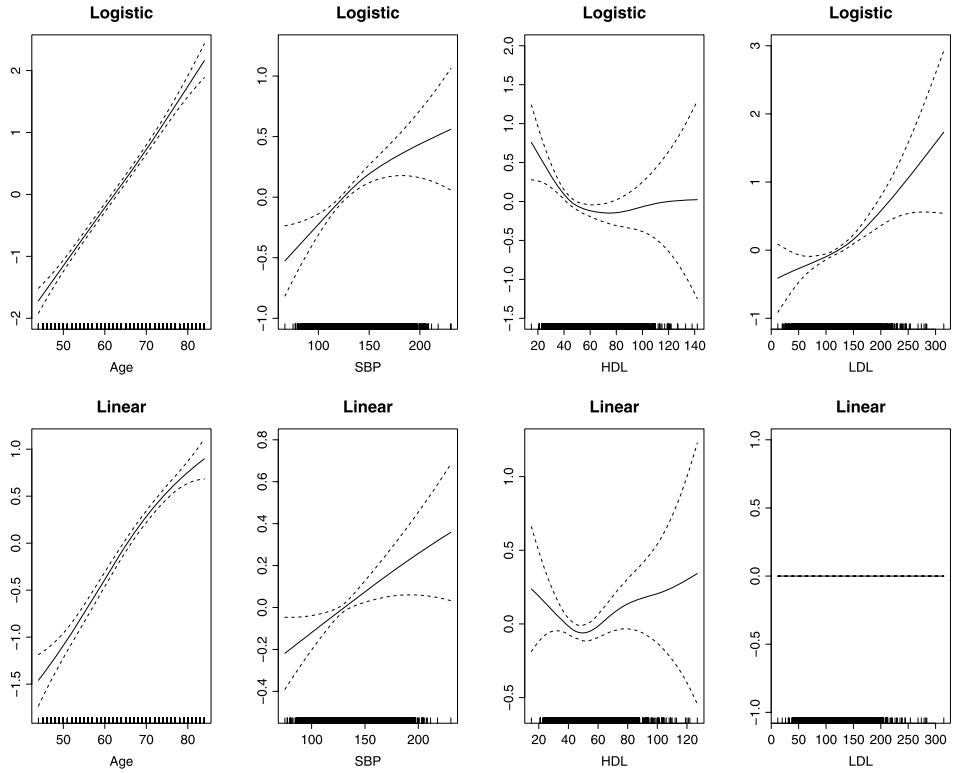}

  \caption{Nonparametric smooth function estimates of the fitted unconstrained zero-inflated
  normal model defined by equations (\protect\ref{mesa:p}) and (\protect\ref{mesa:mu}).
  The dashed lines constitute 95\% point-wise confidence bands. The DBP effects are estimated
  to be zero in both the logistic and linear parts (not shown).}
  \label{fig:mesa}
\end{figure}

The estimated nonparametric smooth functions are displayed in
Figure~\ref{fig:mesa}. Table~\ref{tab:npar} summarizes the
significance test results of the nonparametric terms based on F tests with the
null hypothesis that the smooth function is identically zero over
the observed domain. Age is positively related to both the
probability of positive CAC ($p < 0.001$) and positive CAC level ($p
< 0.001$). However, its effect in the positive mean response $\mu$
shows some curvature at the right tail, suggesting that the age effect is not
as strong in the old as in mid-aged people.
Elevated systolic blood pressure is associated with increased risk of having positive CAC ($p < 0.001$)
and higher CAC score when it is present ($p = 0.003$). Its effect on
the presence of CAC is nonlinear on the logistic scale, whereas it
is almost linear in the positive mean response part. The probability of
having positive CAC decreases as the HDL cholesterol level increases
up to around 60 mg$/$dL, beyond which the risk then stays stable ($p <
0.001$). Among the participants who have positive CAC scores, those
with HDL between 40 to 60 mg$/$dL were observed to have lower CAC levels, however,
its influence is not statistically significant ($p = 0.128$).
LDL is a pronounced risk factor of CAC initiation ($p < 0.001$).
Nevertheless, the LDL effect on the extent and severity of CAC when it
is positive is completely eliminated by the shrinkaged cubic spline. The same was observed
as to the diastolic blood pressure effects in both logistic and linear parts (not shown in Figure~\ref{fig:mesa}).

This study may significantly differ from published MESA studies
concerning the nonparametric covariate effects along the following perspectives.
First, the unconstrained semiparametric zero-inflated model of the Agatston score was found to
have the best prediction performance based on the likelihood cross-validation procedure.
Second, the age effect on the magnitude of CAC when it is positive, and the systolic blood pressure influence
on the probability of having positive CAC, were both observed to be nonlinear.
Third, LDL was shown to have no effect in predicting CAC level among those with positive CAC.
And last, diastolic blood pressure was found not to be a risk factor in human CAC development
by the cubic regression spline with shrinkage adopted in our
study.\vspace*{-3pt}

\begin{table}
\tabcolsep=0pt
\caption{Nonparametric smooth function estimates
 of the fitted unconstrained zero-inflated normal~model defined by equations (\protect\ref{mesa:p}) and (\protect\ref{mesa:mu}).
 EDF stands for effective degrees~of~freedom.~The p-values are based on F tests for significance}\label{tab:npar}
\begin{tabular*}{\textwidth}{@{\extracolsep{\fill}}lcccccc@{}}
  \hline
  & \multicolumn{3}{c}{\textbf{Logistic}} & \multicolumn{3}{c@{}}{\textbf{Linear}} \\[-5pt]
  & \multicolumn{3}{c}{\hrulefill} &
  \multicolumn{3}{c@{}}{\hrulefill}\\
   & \textbf{EDF} & \textbf{F statistic} & \textbf{\emph{p}-value} & \textbf{EDF} & \textbf{F statistic} & \textbf{\emph{p}-value} \\
  \hline
  $s(\mathit{Age})$ & $2.4$ & $816.6$ &  ${<} 0.001$ & $2.7$ & $116.7$ & ${<} 0.001$ \\
  $s(\mathit{DBP})$ & NA & NA & NA & NA & NA &  NA \\
  $s(\mathit{SBP})$ & $1.7$ & \hphantom{8}$31.1$ &  ${<} 0.001$ & $1.0$ & \hphantom{11}$7.2$ &  \hphantom{${<}$}$0.003$ \\
  $s(\mathit{HDL})$ & $3.0$ & \hphantom{8}$19.5$ &  ${<} 0.001$ & $2.8$ & \hphantom{11}$1.8$ &  \hphantom{${<}$}$0.128$ \\
  $s(\mathit{LDL})$ & $2.2$ & \hphantom{8}$38.0$ &  ${<} 0.001$ & NA & NA &  NA \\
  \hline
\end{tabular*}
\end{table}

\section{Discussion and conclusion}
\label{sec:discussion}

We have presented a highly flexible semiparametric regression model for
analyzing zero-inflated data. Possible partial proportional
constraints, whose biological interpretation could be traced to some latent threshold model
under a possibly misspecified link function,
were considered to promote estimation efficiency and help to
reveal the connection between the zero and nonzero data generating
processes. In order to choose the optimal model specification among
multiple candidate models with various partial constraints, a
model selection procedure based on cross-validated likelihood was used,
which was empirically corroborated by a simulation study.
The proposed partially constrained zero-inflated model framework makes it possible
to provide evidence-based justification to address research questions concerning the underlying mechanisms that
drive the presence and magnitude of the nonzero response.
In particular, it can be used to identify closely related covariate effects in the
zero and nonzero data generating processes.
We have adopted the cubic regression spline with shrinkage
to estimate nonparametric smooth functions and select relevant variables simultaneously,
which works well empirically in both simulation and real data application. However,
its theoretical properties still need to be investigated in the future.

When applied to the MESA data analysis, the semiparametric
zero-inflated modeling approach indicates that the initiation of
calcium in the human coronary\vadjust{\goodbreak} artery and the magnitude of positive
calcium (measured by Agatston score) in the general population are
better characterized by an unconstrained zero-inflated model.
It is statistically justified that the initiators of coronary
artery disease may be different from the factors that are related to
extent and progression of the disease which is reflected by the
amount of CAC in those with positive CAC scores. In particular, age
and systolic blood pressure are both risk factors in influencing the development
of CAC from zero to positive, and from small to large amount. But
their effects show some extent of nonlinearity at certain stages. HDL
and LDL cholesterol levels both have pronounced nonlinear effects in
predicting the presence of CAC. However, only HDL has some
impact (not statistically significant) on the extent of CAC in those who have positive CAC scores.
These results may reflect the fact that the biological mechanisms
underlying the initiation and progression of CAC are somehow different.
The partially constrained semiparametric zero-inflated modeling approach (including the
unconstrained case) with the model selection procedure based on likelihood cross-validation
can be applied widely to complex data analysis with the zero-inflation problem.

\section*{Acknowledgments}
We thank the investigators, the staff and the participants of MESA
for their valuable contributions.\footnote{A full list of participating MESA
investigators and institutions can be found at
\url{http://www.mesa-nhlbi.org/}.}
We also thank the Editor, Associate Editor and two reviewers whose comments
resulted in great improvement in the manuscript.


\printaddresses

\end{document}